\def\year{2021}\relax
\documentclass[letterpaper]{article} 
\usepackage{aaai21}  
\usepackage{times}  
\usepackage{helvet} 
\usepackage{courier}  
\usepackage[hyphens]{url}  
\usepackage{graphicx} 
\usepackage{natbib}  
\usepackage{caption} 
\usepackage{tipa}
\usepackage{algorithm}
\usepackage{algorithmic}
\usepackage{amsmath}
\usepackage{graphicx}
\usepackage{multirow}
\usepackage{wrapfig}
\usepackage{color}
\usepackage{booktabs}
\usepackage[switch]{lineno}

\title{UWSpeech: Speech to Speech Translation for Unwritten Languages}

\author{
    Chen Zhang\textsuperscript{\rm 1},
    Xu Tan\textsuperscript{\rm 2},
    Yi Ren\textsuperscript{\rm 1},
    Tao Qin\textsuperscript{\rm 2},
    Kejun Zhang\textsuperscript{\rm 1},
    Tie-Yan Liu\textsuperscript{\rm 2}
}
\affiliations{

    \textsuperscript{\rm 1}Zhejiang University, China \\
    \textsuperscript{\rm 2}Microsoft Research Asia \\
    zc99@zju.edu.cn, xuta@microsoft.com, rayeren@zju.edu.cn,\\
    taoqin@microsoft.com, zhangkejun@zju.edu.cn, tyliu@microsoft.com

}
\begin{document}

\maketitle

\begin{abstract}
Existing speech to speech translation systems heavily rely on the text of target language: they usually translate source language either to target text and then synthesize target speech from text, or directly to target speech with target text for auxiliary training. However, those methods cannot be applied to unwritten target languages, which have no written text or phoneme available. In this paper, we develop a translation system for unwritten languages, named as UWSpeech, which converts target unwritten speech into discrete tokens with a converter, and then translates source-language speech into target discrete tokens with a translator, and finally synthesizes target speech from target discrete tokens with an inverter. We propose a method called XL-VAE, which enhances vector quantized variational autoencoder (VQ-VAE) with cross-lingual (XL) speech recognition, to train the converter and inverter of UWSpeech jointly. Experiments on Fisher Spanish-English conversation translation dataset show that UWSpeech outperforms direct translation and VQ-VAE baseline by about 16 and 10 BLEU points respectively, which demonstrate the advantages and potentials of UWSpeech.
\end{abstract}

\section{Introduction}
Speech to speech translation~\citep{lavie1997janus,nakamura2006atr,wahlster2013verbmobil,jia2019direct} is important to help the understanding of cross-lingual spoken conversations and lectures, and has been used in scenarios such as international travel or conference. Existing speech to speech translation systems either rely on target text as a pivot (they first translate source speech into target text and then synthesize target speech given the translated text~\citep{lavie1997janus,nakamura2006atr,wahlster2013verbmobil}), or directly translate source speech into target speech~\citep{jia2019direct}. In these translation systems, the text corresponding to the target speech is leveraged as either pivots or auxiliary training data~\citep{jia2019direct}; otherwise, the translation would not be possible or the translation accuracy would drop dramatically~\citep{jia2019direct}. 

However, there are thousands of unwritten languages in the world~\citep{lewis2013simons,scharenborg2020speech,godard2017very}, which are purely spoken and have no written text. It is challenging to build speech translation systems for these unwritten languages without text as pivots or auxiliary training data like in~\citet{jia2019direct}. Continuous speech (which usually contains content, context, speaking style, etc.) is much more flexible to represent semantic meanings than discrete symbols (text)~\citep{van2017neural,vigliocco2004representing}, which makes the translation into speech harder than translation into text. Therefore, the key to ease the speech translation for unwritten languages is to reduce the flexible continuous space of speech into a more restricted discrete space.

A variety of previous works~\citep{muthukumar2014automatic,chen2015parallel,wilkinson2016deriving,adams2017automatic,kamper2017embedded,dunbar2017zero,eloff2019unsupervised,tjandra2019vqvae,duong2016attentional,salesky2019exploring} have investigated the conversion between speech and their corresponding phonetic categories (discrete tokens) in an unsupervised manner, which mimics the way that human infants learn acoustic models in their mother tongue during their early years of life~\citep{versteegh2016zero} (some of them only focus on a much easier task such as speech-to-text translation~\citep{duong2016attentional,salesky2019exploring}). Among these works, vector quantized variational autoencoder (VQ-VAE)~\citep{van2017neural,Dunbar_2019,tjandra2019vqvae,chorowski2019unsupervised,liu2019towards,tjandra2019speech,baevski2019vq} has been widely adopted and shown advantages over other methods. However, VQ-VAE is still purely unsupervised and cannot ensure the quality of the learned discrete representations. Therefore, although VQ-VAE performs very well on relatively easier tasks like speech synthesis~\citep{Dunbar_2019}, it cannot achieve good accuracy on more complicated speech to speech translation where semantic representations of speech are important and more accurate phonetic representations are required. Few works tackle on speech to speech translation for unwritten languages~\citep{tjandra2019speech} since it is extremely challenging.

In this paper, we develop UWSpeech (UW is short for UnWritten), a translation system for unwritten languages with three key components: 1) a converter that transforms unwritten target speech into discrete tokens, 2) a translator that translates source-language speech into target-language discrete tokens, and 3) an inverter that converts the translated discrete tokens back to unwritten target speech. As can be seen, the discretization (transform speech into discrete tokens using converter) and reconstruction (synthesize speech from discrete tokens using inverter) steps in UWSpeech is important to ensure translation accuracy.

To this end, we propose XL-VAE, which improves the discretization and reconstruction capability based on VQ-VAE. Different from VQ-VAE that purely relies on unsupervised methods for discrete representation learning, XL-VAE leverages written languages with phonetic labels to improve the vector quantization (discrete representations learning) of unwritten languages through cross-lingual (XL) transfer. As human beings share similar vocal organs and pronunciations~\citep{wind1989evolutionary}, no matter which spoken languages they use, the phonetic representations learned in one language can more or less (depending on the language similarity) help the learning of phonetic representations in another language~\citep{yallop2007introduction,kuhl2008phonetic}. Therefore, XL-VAE can benefit from other written languages and outperform purely unsupervised VQ-VAE on discretizing speech into discrete tokens and synthesizing speech from discrete tokens, and thus enable UWSpeech to achieve better translation accuracy. 

Our contributions can be summarized as follows:
\begin{itemize}
\item We develop UWSpeech, a speech to speech translation system for unwritten languages, and design a novel XL-VAE to train the converter and inverter in UWSpeech jointly for discrete speech representations. 
\item We conduct experiments on Fisher Spanish-English speech conversation dataset, assuming the target language is unwritten. Experiment results show that UWSpeech equipped with XL-VAE achieves 16 and 10 BLEU points improvements over direct translation and VQ-VAE baseline respectively, which demonstrates the advantages and potentials of UWSpeech on speech to speech translation for unwritten target languages. \footnote{Speech samples and experimental details can be found in https://speechresearch.github.io/uwspeech/}
\item We further apply UWSpeech to text to speech translation and speech to text translation for unwritten languages. The improvements over direct translation and VQ-VAE baseline demonstrate the general applicability of UWSpeech beyond speech to speech translation.
\end{itemize}

\section{Background}
\paragraph{A Taxonomy of Speech Translation and Our Focused Setting}
Based on the successes of text to text translation~\citep{bahdanau2014neural,luong2015effective,vaswani2017attention}, speech translation~\citep{berard2016listen,weiss2017sequence,jia2019direct} has been developed to handle speech as translation input and/or output. Previous works on speech translations has evolved from cascaded models~\citep{ney1999speech,matusov2005integration,lavie1997janus,nakamura2006atr,wahlster2013verbmobil} to end-to-end models~\citep{berard2016listen,weiss2017sequence,vila2018end,sperber2019attention,jia2019direct}, where the text corresponding to speech is leveraged as auxiliary training~\citep{jia2019direct} for better accuracy. Depending on the speech is in the source or/and target side, speech translation can be divided into three categories: speech to text translation, text to speech translation and speech to speech translation. In this paper, we focus on the most difficult setting: speech to speech translation for unwritten languages. In this way, we can not leverage any source or target text in auxiliary tasks like in~\citet{jia2019direct}. Furthermore, we also extend UWSpeech for text to speech translation with unwritten target languages and speech to text translation with unwritten source languages to demonstrate the generalization ability of our method. 
Besides, our method can also be applied to the written target languages whose text or phonetic transcripts are not available in the training data.

\paragraph{Discrete Speech Representations}
Learning discrete representations of speech has long been studied for better speech understanding and modeling. Previous works on discrete speech representations include k-means clustering~\citep{kamper2017embedded,dunbar2017zero}, Gaussian mixture model clustering~\citep{chen2015parallel}, tree-based clustering~\citep{muthukumar2014automatic}, binarization with straight-through estimation~\citep{eloff2019unsupervised}, categorical VAE~\citep{eloff2019unsupervised} and the more advanced vector quantized
VAE (VQ-VAE)~\citep{van2017neural,Dunbar_2019,tjandra2019vqvae,chorowski2019unsupervised,liu2019towards,tjandra2019speech,baevski2019vq}. VQ-VAE has been widely used to cluster/quantize the representations of speech and discretize into codebook sequence, and has achieved good results on some tasks such as subword units discovery from speech or text to speech synthesis~\citep{Dunbar_2019}. However, VQ-VAE is a purely unsupervised clustering method for discrete speech representations, which limits its effectiveness on harder tasks like speech translation. In this paper, we improve VQ-VAE with cross-lingual (XL) speech recognition and propose XL-VAE to achieve better discrete speech representations.

\section{UWSpeech}
In this section, we introduce the design of our proposed UWSpeech: a speech to speech translation system for unwritten target languages with the help of cross-lingual vector quantized variational autoencoder (XL-VAE). We first describe the overall pipeline of UWSpeech, and then introduce the detailed design of XL-VAE.

\begin{figure}[!thb]
	\centering
	\includegraphics[width=0.48\textwidth,trim={8cm 8cm 9cm 7.8cm}, clip=true]{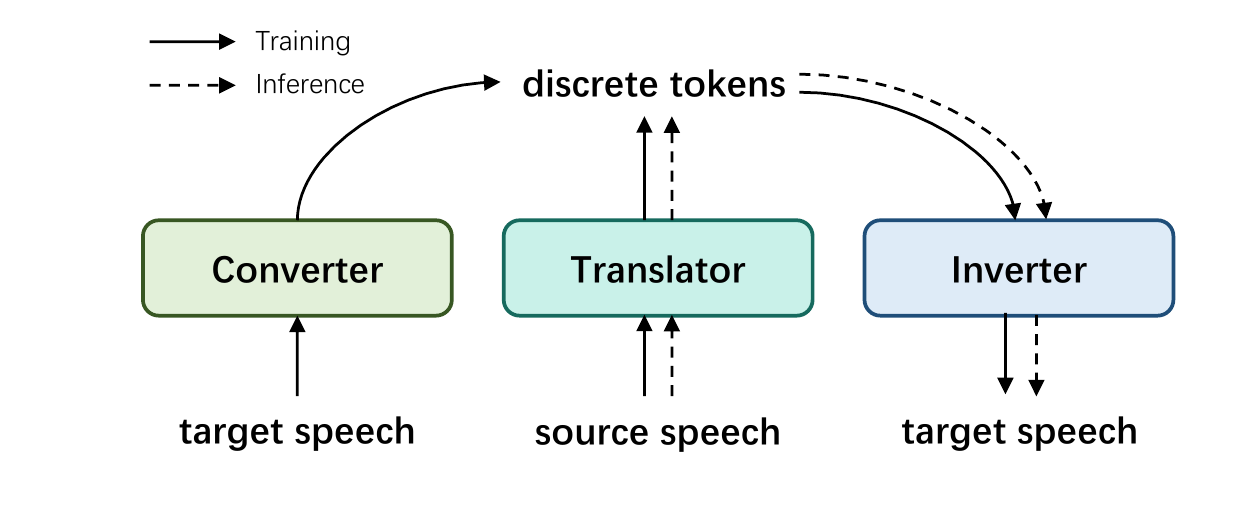}
	\vspace{-0.8cm}
	\caption{The training and inference pipeline of UWSpeech.}
	\label{UWSpeech_pipeline}
\end{figure}

\subsection{Pipeline Overview}
For speech to speech translation where the target language is unwritten, UWSpeech consists of three components as shown in Figure~\ref{UWSpeech_pipeline}: 1) a converter to transform the target-language speech into discrete tokens; 2) a translator to translate the source speech into target discrete tokens; 3) an inverter to convert the target discrete tokens back to target speech. We introduce each component in the following subsections.

\paragraph{Translator} 
Denote the training corpus as $\{ (x, y) \in (\mathcal{X}, \mathcal{Y})\}$, where $x$ and $y$ are the source and target speech sequence. According to the pipeline of UWSpeech, we convert the target unwritten speech sequence $y \in \mathcal{Y} $ into discrete token sequence $z \in \mathcal{Z}$ to form a triple corpus $(\mathcal{X}, \mathcal{Z}, \mathcal{Y})$. We train a machine translator $\theta_{\text{trans}}$ by minimizing the negative log-likelihood loss
\begin{equation}
\small
\mathcal{L}_{\text{trans}} = -\sum_{(x,z)\in (\mathcal{X}, \mathcal{Z})} \log P(z|x;\theta_{\text{trans}}),
\label{eq_mt_loss}
\end{equation}
where $\theta_{\text{trans}}$ can be implemented as a standard encoder-attention-decoder~\citep{vaswani2017attention} based model with several convolution layers in the encoder to handle speech input, and will be described in the experiment setting.

\paragraph{Converter and Inverter}
The converter and inverter transform the speech sequence $y$ into discrete token sequence $z$ and transform $z$ back to speech sequence $y$ respectively, and follow the form of autoencoder where the converter acts like the encoder and the inverter acts like the decoder. Inspired by VQ-VAE, we propose a novel XL-VAE to better train the converter and inverter for speech translation.

\subsection{XL-VAE}
\begin{figure}[!thb]
	\centering
	\includegraphics[width=0.5\textwidth,trim={10cm 4.4cm 9cm 4cm}, clip=true]{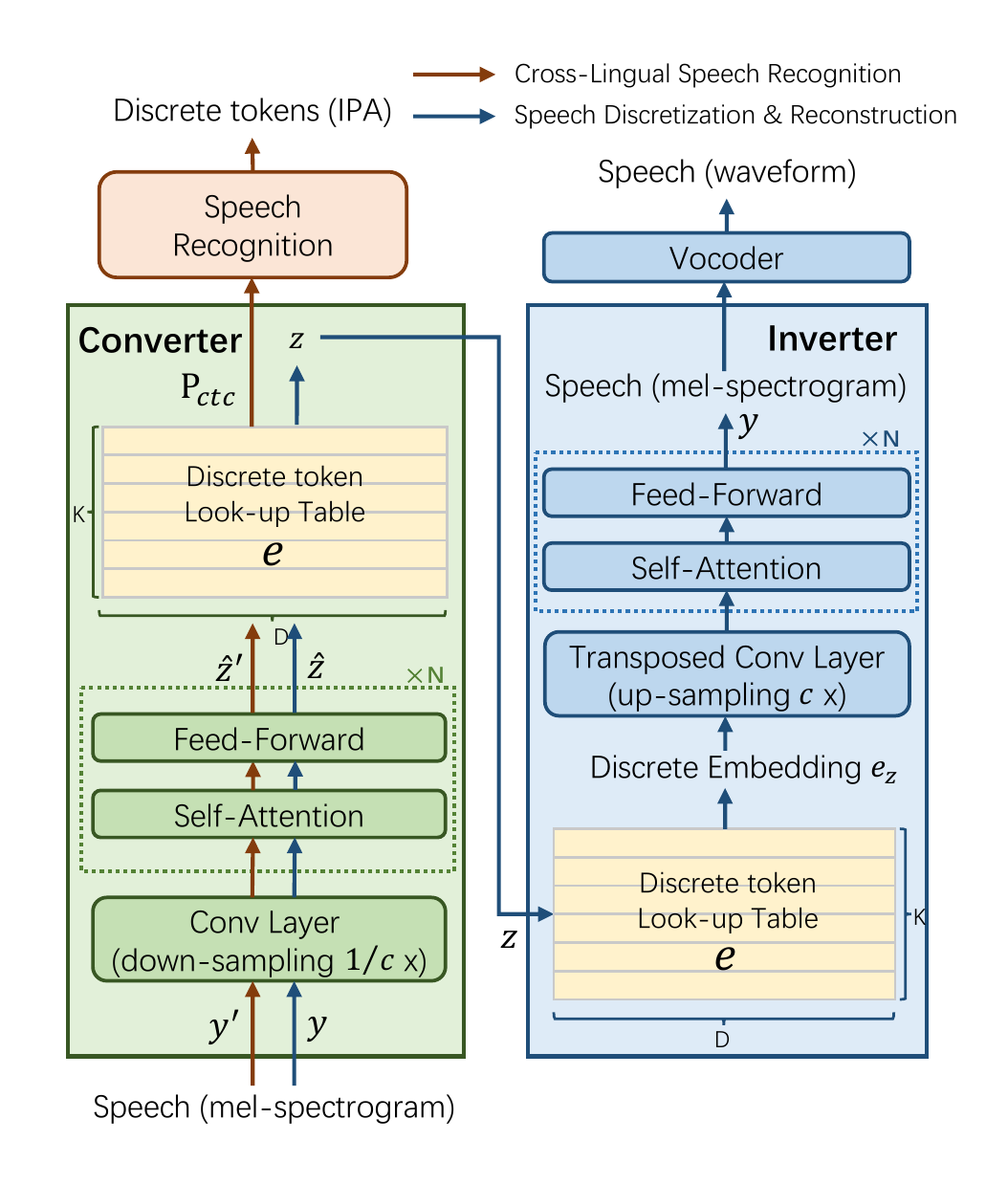}
	\vspace{-0.4cm}
	\caption{The model structure of XL-VAE.}
	\label{fig_xl_vae}
\end{figure}

XL-VAE first encodes the speech sequence into hidden representations to extract discrete tokens with a converter, and reconstructs the original speech sequence given the representations of discrete tokens with an inverter. Different from VQ-VAE~\citep{van2017neural}, XL-VAE extracts discrete representations not by unsupervised vector clustering, but by speech/phoneme recognition, where the recognition capability is transferred from other popular written languages. We train the phoneme recognition on written languages with speech and phoneme pairs based on the converter. We illustrate XL-VAE in Figure~\ref{fig_xl_vae} and formulate each module in XL-VAE as follows.

\paragraph{Converter} The converter of XL-VAE $\theta_{\text{conv}}$ takes speech sequence $y$ as input and generate continuous hidden representations $\hat{z}$: 
\begin{equation}
\small
\begin{aligned}
\hat{z} &= f(y; \theta_{\text{conv}}). 
\end{aligned}
\label{eq_xlvae_enc}
\end{equation}
$\hat{z}$ is further converted into discrete latent variables $z$ through nearest neighbour search based on dot-product\footnote{We use dot-product here instead of Euclidean distance in VQ-VAE, in order to be consistent with  the speech recognition where the hidden representations $\hat{z}'$ are multiplied with the matrix $e$ and then transformed through a softmax function to get the probability of each phoneme category (which is described in the later part of this subsection).}: 
\begin{equation}
\small
q(z = k|y) =  \left\{
             \begin{array}{lr}
             1~~\text{for}~k = \arg\max_{i}~ (\hat{z} * e_i) \\
             0~~\text{otherwise}
             \end{array},
\right. 
\label{eq_xlvae_lookup}
\end{equation}
where $q(z|y)$ denotes the categorical distribution of the discrete variable $z$.  $e \in \mathcal{R}^{K\times D}$ denotes the embedding space of the discrete tokens, $K$ denotes the number of discrete tokens and D denotes the size of each embedding vector $e_i$ for $i \in \{1, 2,...,K\}$. 

As shown in Figure~\ref{fig_xl_vae}, the converter takes speech (mel-spectrogram) sequence as input and uses several convolution layers with strides to reduce the length of speech sequence by $1/c$. It then stacks $N$ Transformer blocks~\citep{vaswani2017attention}, where each block contains a self-attention layer and a feed-forward layer with a layer-normalization and a residual connection on top of each layer. For a speech sequence with length of $l$, the generated discrete tokens $z$ has length of $l/c$.

\paragraph{Inverter} The inverter of XL-VAE $\theta_{\text{inv}}$ takes discrete tokens $z$ as input and convert $z$ into $e_z$ with discrete token look-up table $e$ (the same $e$ as used in the converter). Then $e_z$ is used to reconstruct the original speech sequence $y$: 
\begin{equation}
\small
\begin{aligned}
\mathcal{L}_{\text{inv}} = \sum_{y \in \mathcal{Y}}  (y - f(e_z; \theta_{\text{inv}}))^2.
\end{aligned}
\label{eq_xlvae_dec}
\end{equation}

As shown in Figure~\ref{fig_xl_vae}, the inverter leverages several transposed convolution layers~\citep{dumoulin2016guide} to increase the length of $e_z$ by $c\times$ (opposed to the $1/c\times$ in the converter), to match the length of the original mel-spectrogram sequence. It then stacks $N$ Transformer blocks~\citep{vaswani2017attention} as used in the converter. The inverter reconstructs the speech sequence in parallel. Therefore, different from the conventional self-attention in Transformer decoder which cannot see the information in the future positions, the self-attention in the inverter can see the information in all positions, just like the converter. A vocoder~\citep{griffin1984signal,oord2016wavenet} is leveraged to further convert the mel-spectrogram into an audio waveform.

\paragraph{Cross-Lingual (XL) Speech Recognition} Instead of unsupervised quantization in VQ-VAE, XL-VAE introduces speech recognition in other written languages to help learn the discrete representations, as shown in Figure~\ref{fig_xl_vae}. Given the speech and phoneme sequence pairs $(y', t') \in (\mathcal{Y}', \mathcal{T}')$ of written languages, we use the converter $\theta_{\text{conv}}$ to transform speech $y'$ into $\hat{z}'$, and then multiply $\hat{z}'$ with the discrete token embedding matrix $e$ ($e$ is denoted in Equation~\ref{eq_xlvae_lookup}) and get the probability distribution $P_{\text{ctc}}$ over $K$ phoneme categories with a softmax operation, where $K$ is size of phoneme vocabulary in the written languages, and also the number of discrete tokens in $e$, which is similar with ~\citet{li2020towards}. We train the phoneme recognition with connectionist temporal classification (CTC) loss~\citep{graves2006connectionist}. The formulation of the cross-lingual speech recognition is as follows:
\begin{equation}
\small
\begin{aligned}
    \hat{z}' &= f(y'; \theta_{\text{conv}}), 
    ~~~~~P_{\text{ctc}}(r) = \prod_{i=1}^{|r|}\text{softmax} ( \hat{z}' * e )_{r_{i}}, \\
    \mathcal{L}_{\text{xl}} &= -\sum_{(x', t') \in (\mathcal{Y}',\mathcal{T}')} \sum_{s \in \phi(t')} \log P_{\text{ctc}}(r=s),
\end{aligned}
\label{eq_xlvae_xl}
\end{equation}
where $\phi(t')$ denotes the set of valid CTC paths for phoneme sequence $t'$, $P_{\text{ctc}}(r=s)$ denotes the probability of the CTC path $s$, $\text{softmax}(\cdot)_{r_{i}}$ denotes the probability of observing label $r_{i}$ under the softmax function and $|r|$ denotes the length of sequence $r$. The loss function $\mathcal{L}_{\text{xl}}$ aims to minimize the negative log-likelihood of all the valid CTC paths in the training set. For more details of CTC, you can refer to~\citet{graves2006connectionist}, which is not the focus of this work. 

\paragraph{Discrete Representation} We choose international phonetic alphabet (IPA)~\citep{international1999handbook} as the phoneme set of the written languages. In this way, the discrete token embeddings $e \in \mathcal{R}^{K\times D}$ are exactly the embeddings of IPA where $K$ is the size of IPA set and $D$ is the dimension of the embedding vector. The unwritten speech is converted into discrete tokens which fall into the IPA set of written languages. The discrete tokens $z$ as well as the corresponding embedding vectors in $e$ are taken as the discrete representations of speech $y$.

\paragraph{Loss Function of XL-VAE}
Putting Equation~\ref{eq_xlvae_enc},~\ref{eq_xlvae_lookup},~\ref{eq_xlvae_dec} and~\ref{eq_xlvae_xl} together, we have the loss function of XL-VAE:
\begin{equation}
    \mathcal{L}_{\text{xl-vae}} = \mathcal{L}_{\text{inv}} + \lambda \mathcal{L}_{\text{xl}},
\label{eq_xlvae_loss}
\end{equation}
where $\lambda$ is a hyperparameter to trade-off the two loss terms.

\subsection{Training and Inference}
Finally, we describe the training and inference procedure of UWSpeech according to the formulations in the previous two subsections. The detailed procedure is shown in Algorithm~\ref{alg_UWSpeech}. 
\begin{algorithm}[htb]
\small
\caption{UWSpeech Training and Inference}
\label{alg_UWSpeech}
\begin{algorithmic}
    \STATE {\bfseries Training:}
    \STATE {\bfseries Input:} Speech to speech translation corpus $(\mathcal{X}, \mathcal{Y})$ where $\mathcal{Y}$ represents target unwritten speech. Paired speech and phoneme corpus $(\mathcal{Y}', \mathcal{T}')$ in written languages where $\mathcal{T}'$ uses IPA as the phoneme set.
    \STATE {\bfseries Step 1:} Train the XL-VAE model with corpus $\mathcal{Y}$ and $(\mathcal{Y}', \mathcal{T}')$ using loss in Equation~\ref{eq_xlvae_loss} to obtain the converter $\theta_{\text{conv}}$, inverter $\theta_{\text{inv}}$ and discrete token look-up table $e$. 
    \STATE {\bfseries Step 2:} Convert the unwritten speech corpus $\mathcal{Y}$ into discrete sequence corpus $\mathcal{Z}$ following Equation~\ref{eq_xlvae_enc} and~\ref{eq_xlvae_lookup}. Train the machine translator $\theta_{\text{trans}}$ with corpus $(\mathcal{X}, \mathcal{Z})$ using loss in Equation~\ref{eq_mt_loss}.
\\\hrulefill 
   \STATE {\bfseries Inference:}
   \STATE {\bfseries Input:} Source speech corpus $\mathcal{X}$, translator $\theta_{\text{trans}}$, discrete token look-up table $e$ and inverter $\theta_{\text{inv}}$.
   \STATE {\bfseries Step 1:} For each speech sequence $x \in \mathcal{X}$, generate target  discrete tokens: $z\sim P(z|x; \theta_{\text{trans}})$.
   \STATE {\bfseries Step 2:} Convert $z$ into $e_z$ through discrete token look-up table $e$, and synthesize target speech: $y = f(e_z; \theta_{\text{inv}})$.
\end{algorithmic}
\end{algorithm}
\vspace{-0.5cm}

\section{Experiments and Results}
In this section, we first introduce the experimental setup and then report the results of UWSpeech for speech to speech translation. We further conduct some analyses of UWSpeech. Finally, we also apply UWSpeech to text to speech translation and speech to text translation settings.

\subsection{Experimental Setup}
\paragraph{Datasets}
\label{datasets}
Following the common practice in low-resource and unsupervised speech and translation works~\citep{lample2018unsupervised,song2019mass,ren2019almost}, we conduct experiments on popular written languages but remove the text of target speech to simulate unwritten languages. We choose Fisher Spanish-English dataset~\citep{post2013improved} for translation. Considering 1) translation to unwritten languages is difficult and 2) the most useful translation scenarios for unwritten languages are daily communication, travel translation, etc., where high-frequency and simple words/sentences are usually used, we choose some common sentences from the original full test set to form our test set \textcolor{black}{(denoted as common test set)}. 
But we still show the results on the full test set of the main experiments setting in Table~\ref{tab_s2s_esen} and Table~\ref{tab_s2s_enes} for reference. 
For the written languages used in XL-VAE, we choose French, German and Chinese with speech data and corresponding phoneme sequence. 
Both the German and French datasets are from Common Voice\footnote{https://voice.mozilla.org/}, where the German corpus contains about 280K training examples (325 hours) with 5007 different speakers and the French corpus contains 150K training examples (173 hours) with 3005 different speakers. For the Chinese dataset, we use AIShell~\citep{aishell_2017} which contains about 140K training examples (178 hours) with 400 different speakers.

\paragraph{Model Configuration}
We choose Transformer~\citep{vaswani2017attention} as the basic model structure for the converter, inverter and translator, since it achieves good results on machine translation, speech recognition and speech synthesis tasks. 

\paragraph{Training Details}
We first train the converter, inverter and discrete token embeddings in XL-VAE. We up-sample the speech data of each written language (German, French, Chinese) to the same amount, and then up-sample the speech data of unwritten language (English or Spanish) to match the total amount of written languages. We ensure there are an equal amount of data in written and unwritten languages in each mini-batch. We choose the $\lambda$ in Equation~\ref{eq_xlvae_loss} according to the validation performance and set $\lambda$ to $0.01$. The batch size is set to $25$K frames for each GPU and the XL-VAE training takes $200$K steps on $4$ Tesla V100 GPUs. 

After the training of XL-VAE, the phoneme error rates (PER) of three written languages (German, French and Chinese) on the development set are $16\%$, $21\%$ and $12\%$ respectively. We convert the target unwritten speech into the discrete token sequence and keep the output discrete token sequence as it is, without removing any special or repeated tokens. We use the discrete token sequence generated by XL-VAE to train translator, with a batch size of $16$K frames on each GPU and $100$K training steps on $4$ Tesla V100 GPUs. 

Our code is implemented based on tensor2tensor library~\citep{vaswani2018tensor2tensor}\footnote{https://github.com/tensorflow/tensor2tensor}. 

\paragraph{Inference and Evaluation}
During inference, we use the translator to generate discrete token sequences from source speech with beam search. We set beam size to $4$ and the length penalty to $1.0$. We then directly use the inverter to transform the discrete token sequence back to target speech. 

To evaluate the accuracy of the speech translation, following the practice in~\citet{jia2019direct}, we pre-train an automatic speech recognition model (which can achieve 85.62 BLEU points on our test set and is comparable with~\citet{jia2019direct}) to generate the corresponding text of the translated speech, and then calculate the BLEU score~\citep{papineni2002bleu} between the generated text and the reference text. We report BLEU score using case insensitive BLEU with moses tokenizer\footnote{https://github.com/moses-smt/mosesdecoder/blob/master/ scripts/tokenizer/tokenizer.perl} and multi-bleu.perl\footnote{https://github.com/moses-smt/mosesdecoder/blob/master/ scripts/generic/multi-bleu.perl}. Due to the Fisher corpus has $4$ English references in the test set, we report 4-reference BLEU score for Spanish to English setting, and still report single-reference BLEU score for English to Spanish setting.

\subsection{Results}
In this subsection, we report the experiment results of UWSpeech. We compare UWSpeech mainly with two baselines: 1) \textit{Direct Translation}, which directly translates the source speech into target speech in an encoder-attention-decoder model without any text as auxiliary training data or pivots. 

2) Discretization with VQ-VAE (denoted as \textit{VQ-VAE}), which follows the translation pipeline in UWSpeech but replaces XL-VAE with original VQ-VAE for speech discretization.

\begin{table}[!h]
\centering
\small 
\begin{tabular}{c | c c c }
\toprule
Method & \textit{Direct Translation} & \textit{VQ-VAE} & \textit{UWSpeech} \\
\midrule
common test set & 1.45 & 7.17 & 17.33 \\
full test set & 0.8 & 3.42 & 9.35 \\
\bottomrule
\end{tabular}
\vspace{-0.2cm}
\caption{The BLEU scores of Spanish to English speech to speech translation, where English is taken as the unwritten language.}
\label{tab_s2s_esen}
\vspace{-0.4cm}
\end{table}

The speech to speech translation results on Spanish to English are shown in Table~\ref{tab_s2s_esen}. As can be seen, \textit{Direct Translation} achieves a very low BLEU score, which is consistent with the findings in \citet{jia2019direct} and demonstrates the difficulty of direct speech to speech translation. \textit{VQ-VAE} achieves slightly better BLEU score than \textit{Direct Translation}, but still with poor accuracy, which demonstrates the limitations of the purely unsupervised method for speech discretization when handling speech translation. \textcolor{black}{On the common test set as we described in Section~\ref{datasets}, }\textit{UWSpeech} achieves 17.33 BLEU points, about 10 points higher than \textit{VQ-VAE} and 16 points higher than \textit{Direct Translation}. \textcolor{black}{UWSpeech also shows a huge gain on the full test set.} We also find that the inverter in XL-VAE can get a lower reconstruction loss than VQ-VAE on the validation set, demonstrating that the discrete tokens extracted by XL-VAE can not only help the discrete token translation in translator but can also benefit the speech reconstruction in inverter, which together contributes to the better accuracy in speech translation. The above results demonstrate the advantages of XL-VAE in leveraging cross-lingual speech recognition for speech discretization and the effectiveness of UWSpeech for unwritten speech translation. 

We further show the experiment results on English to Spanish translation in Table~\ref{tab_s2s_enes}. Similar to the results on Spanish to English translation, \textit{Direct Translation} achieves a very low BLEU score and \textit{UWSpeech} achieves about 8 points higher than \textit{VQ-VAE} \textcolor{black}{on the common test set and 7 points higher on the full test set,} demonstrating the effectiveness of UWSpeech.

\begin{table}[!h]
\centering
\small 
\begin{tabular}{c | c c c }
\toprule
Method & \textit{Direct Translation} & \textit{VQ-VAE} & \textit{UWSpeech} \\
\midrule
common test set & 0.80 & 3.12 & 11.13 \\
full test set & 0.62 & 1.45 & 8.27 \\
\bottomrule
\end{tabular}
\vspace{-0.2cm}
\caption{The BLEU scores of English to Spanish speech to speech translation, where Spanish is taken as the unwritten language.}
\label{tab_s2s_enes}
\vspace{-0.6cm}
\end{table}

\subsection{Method Analyses}
\label{method_analysis}
In this subsection, we conduct some experimental analyses on the proposed UWSpeech. For simplicity, we only show the results on the common test set we described in Section~\ref{datasets}.

\paragraph{UWSpeech with Multi-task Training} ~\citet{jia2019direct} proposes a direct speech to speech translation model, which improves translation accuracy through multi-task training (source speech to source text (automatic speech recognition), and source speech to target text (speech to text translation)). Originally, due to lack of text in both source and target languages, speech to speech translation for unwritten languages could not take advantage of the multi-task training mechanism. However, our proposed XL-VAE can discretize the speech into discrete tokens, which can be regarded as text for multi-task training Therefore, we study how  UWSpeech performs when combining with multi-task training. 

We combine UWSpeech with multi-task training in two ways:
\begin{itemize}
    \item SL ASR (Source Language ASR): Training a model that has a shared speech encoder and two decoders: one is for speech recognition on source unwritten languages  (source speech to the corresponding discrete tokens), and the other is for speech translation on source unwritten languages  (source speech to the discrete tokens in the target language). Both of the discrete tokens corresponding to the source and target unwritten languages are generated by XL-VAE. In this way, we leverage automatic speech recognition of source unwritten language (discrete token sequences as target) as auxiliary loss in our Translator.
    \item WL ASR (Written Languages ASR): Training a model that has a shared speech encoder and two decoders: one is for phone-level automatic speech recognition on auxiliary written languages (e.g., German, French, and Chinese in this paper), and the other is for speech to speech translation on unwritten languages  (e.g., translate Spanish speech to English speech directly) at the same time, hoping that ASR can help the speech encoder training better.
\end{itemize}

As we can see in Table~\ref{tab_multitask}, the SL ASR setting can only improve slightly from 17.33 to 17.41, which also demonstrates the discretization of source speech is not so necessary. The BLEU score of the WL ASR setting is very low (2.36), which indicates that the Direct Translation model cannot make full use of the written languages, while XL-VAE can do this well. 

\begin{table}[!h]
\centering
\small 
\begin{tabular}{c | c c c }
\toprule
Method & \textit{UWSpeech} & \textit{SL ASR} & \textit{WL ASR} \\
\midrule
BLEU & 17.33 & 17.41 & 2.36 \\
\bottomrule
\end{tabular}
\caption{The BLEU scores of Spanish to English speech to speech translation, combines with multi-task training in different ways.}
\label{tab_multitask}
\vspace{-0.8cm}
\end{table}

\paragraph{Analyses of Written Languages in XL-VAE} We study the influence of written languages in XL-VAE on the translation accuracy, mainly from two perspectives: 1) the data amount of the written languages, and 2) the similarity between the written and unwritten languages. To this end, we design several different experimental settings for this study, as shown in Table~\ref{tab_anaysis_written_lan}\footnote{Someone may wonder the acoustic conditions of the speech in different written languages may influence the comparison. We listened and compared the acoustic conditions in their speech data and only found little difference. Therefore, we can focus more on the data amount and language similarity instead of acoustic conditions considering the good robustness of the ASR model.}. 

\vspace{-0.2cm}
\begin{table}[!h]
\centering
\small 
\begin{tabular}{l l l }
\toprule
Setting & Configuration & BLEU \\ 
\midrule
\#1 & De (80h)  & 10.58  \\
\#2 & De (160h) & 12.12 \\
\#3 & De (320h) & 15.20 \\ 
\#4 & De (320h) + Fr (160h) + Zh (160h) & 17.33 \\
\#5 & Fr (160h) & 11.79 \\
\#6 & Zh (160h) & 9.38 \\
\bottomrule
\end{tabular}
\vspace{-0.2cm}
\caption{The BLEU scores of Spanish to English speech to speech translation with different written languages as well as different data amounts for XL-VAE. We denote German as De, French as Fr and Chinese as Zh.}
\label{tab_anaysis_written_lan}
\vspace{-0.4cm}
\end{table}

From setting \#1, \#2 and \#3, it can be seen that increasing the data amount of written language (German) can improve the speech translation accuracy. Comparing setting \#4 with \#3, we can find that further adding other languages (French and Chinese) to increase the total data amount can also improve translation accuracy. Comparing setting \#2, \#5 and \#6, we can find that German helps more on the discretization of English than French, and both German and French help more than Chinese, which is consistent with the language similarity. According to the language family~\citep{lewis2013simons}, German and English belong to the same Germanic branch in the Indo-European family, while French and English belong to the same Indo-European family although not in the same branch. Chinese and English belong to different families and are far apart from each other. Even using distant Chinese as written language, our method still achieves higher accuracy than VQ-VAE (9.38 vs 7.17). 

\paragraph{Varying Embedding Size $D$ and Down-Sampling Ratio $c$ in XL-VAE} We further evaluate how the discrete token embedding size $D$ and the speech down-sampling ratio $c$ in XL-VAE influence the translation accuracy. We set $c=4$ when varying $D$ and set $D=256$ when varying $c$ according to preliminary experiments. As shown in Table~\ref{table_emb_size}, discrete token embedding size $D=256$ performs better and down-sampling ratio $c=4$ performs better.

\vspace{-0.3cm}
\begin{table}[!h]
\centering
\small 
\begin{tabular}{ l c c c c }
\toprule
    Embedding Size $D$ & 64 & 128 & 256 & 512 \\ 
    \midrule
    BLEU  & 13.85 & 15.20 & 17.33 & 17.13  \\ 
    \midrule
    \midrule
    Down-Sampling Ratio $c$ & 1 & 2 & 4 & 8 \\ \midrule
    BLEU & 10.05 & 13.27 & 17.33 & 16.85 \\ 
\bottomrule
\end{tabular}
\vspace{-0.2cm}
\caption{The BLEU scores of Spanish to English translation with different discrete token embedding sizes and down-sampling ratios.}
\label{table_emb_size}
\vspace{-0.8cm}
\end{table}

\paragraph{The Advantage of Training Converter and Inventer Jointly} To study the benefits of joint training the converter and inverter in XL-VAE, we separately train the converter by speech recognition on written languages and the inverter by reconstructing speech from discrete tokens. Separate training achieves 13.51 BLEU points on Spanish to English translation, which is much lower than joint training in XL-VAE (17.33), demonstrating the effectiveness of joint training in XL-VAE. Also, the setting points out that even if we pre-train VQ-VAE with the same unwritten language data, it underperforms our UWSpeech. 

\paragraph{Discretization of Source Speech} 
To study the translation accuracy if we also discretize the source speech into discrete tokens at the same time, we conduct experiments on Spanish to English translation direction and achieve 17.45 BLEU points, which is just slightly better than only discretizing target speech (17.33). The results demonstrate that direct translation from source speech is not as difficult as direct translation into target speech.

\paragraph{Case Analyses} We further analyze some translation cases by our UWSpeech system and the baseline methods on Spanish to English translation. As shown in Table~\ref{table_case}, we list the source (Spanish) and target (English) reference text corresponding to the speech, and convert the translated English speech into text with the pre-trained automatic speech recognition model as used in the evaluation. For the first case, both \textit{Direct Translation} and \textit{VQ-VAE} miss the meaning of ``what she said'' while \textit{UWSpeech} can translate the meaning. For the second case, only \textit{UWSpeech} can translate the meaning of ``How's it going, where are you from?'' correctly. We also show the translated discrete token sequence (IPA) by the translator (denoted as \textit{IPA (UWSpeech)}) as well as the discrete token sequence extracted from the target speech (denoted as \textit{IPA (Target)}) in Table~\ref{table_case}. It can be seen that the IPA translated by UWSpeech is close to the target IPA, and both are close to the pronunciation of English speech, which demonstrates the good accuracy of the IPA extracted by XL-VAE and translated by the translator. 
We attach the corresponding speech and more cases at https://speechresearch.github.io/uwspeech.

\begin{table}[!h]
\centering
\small 
\begin{tabular}{ l l }
\toprule
    Spanish (Source) & Yo no entendí lo que ella dijo. \\ 
    English (Target) & I didn't understand what she said. \\
    \midrule
    \textit{Direct Translation} & I don't know. \\
    \textit{VQ-VAE} & I didn't understand. \\
    \textit{UWSpeech} & I didn't understand what she say. \\
    \midrule
    \multirow{2}{*}{\textit{IPA (Target)}} & \textipa{ai ai n | d I g n n z E | 5 n Y s t t @ l 5 n n } \\
    &  \textipa{ t t | v O t | t i: s | E E: n } \\
    \multirow{2}{*}{\textit{IPA (UWSpeech)}} & \textipa{ai ai | d e n n n | a n n v y: s s t e n n n t } \\
    & \textipa{| v O 5 t | t d i: | E E l } \\
\toprule
\bottomrule
    Spanish (Source) & Qué tal, ¿de dónde eres? \\ 
    English (Target) & How's it going, where are you from? \\
    \midrule
    \textit{Direct Translation} & Had a price. \\
    \textit{VQ-VAE} & Like are you there are you from? \\
    \textit{UWSpeech} & How are you, where are you from? \\
    \midrule
    \textit{IPA (Target)} & \textipa{h h a: s | b I t t | Oy n n | O | K | j | v a: K m} \\
    \textit{IPA (UWSpeech)} & \textipa{h h au | 5 | j ø: | 5 | j e | v a: K m} \\
\bottomrule
\end{tabular}
\caption{Some translation cases in Spanish to English speech to speech translation.}
\vspace{-0.4cm}
\label{table_case}
\end{table}

\subsection{Extension of UWSpeech}
Although UWSpeech is designed for speech to speech translation, it can also be applied to other two speech translation settings for unwritten languages: text to speech translation and speech to text translation. We conduct experiments on these two settings on Spanish to English translation to verify the broad applicability of UWSpeech for unwritten speech translation, and show the results \textcolor{black}{on the common test set 
we described in Section~\ref{datasets}} in Table~\ref{tab_s2t_t2s_esen}. 

In the text to speech setting, \textit{Direct Translation} still achieves very poor translation accuracy and \textit{UWSpeech} achieves about 14 BLEU points improvements over \textit{VQ-VAE} baseline, demonstrating the effectiveness of UWSpeech on text to speech translation for unwritten languages.

In the speech to text setting, \textit{UWSpeech} achieves much higher accuracy than \textit{VQ-VAE} and slightly better accuracy than \textit{Direct Translation}. While verifying the effectiveness of our UWSpeech, these results also demonstrate that it is not that necessary to discretize the source speech in speech translation, which is consistent with our findings in Section~\ref{method_analysis}, and is also consistent with the results in~\citet{weiss2017sequence} where even leveraging the ground-truth text corresponding the source speech can only achieve a BLEU gain less than 2 points.

\begin{table}[!ht]
\centering
\small 
\begin{tabular}{c | c c c }
\toprule
Method & \textit{Direct Translation} & \textit{VQ-VAE} & \textit{UWSpeech} \\
\midrule
Text to Speech & 5.47 & 8.02 & 22.03 \\
Speech to Text & 33.87 & 29.98 & 34.05 \\
\bottomrule
\end{tabular}
\vspace{-0.2cm}
\caption{The BLEU scores of the text to speech and speech to text setting on Spanish to English translation, where English and Spanish is taken as the unwritten language in the text to speech setting and speech to text setting respectively.}
\label{tab_s2t_t2s_esen}
\end{table}
\vspace{-0.6cm}

\section{Conclusion}
In this paper, we developed UWSpeech, a speech to speech translation system for unwritten target languages, and designed XL-VAE, an enhanced version of VQ-VAE based on cross-lingual speech recognition, to jointly train the converter and inverter to discretize and reconstruct the unwritten speech in UWSpeech. Experiments on Fisher Spanish-English dataset demonstrate that UWSpeech equipped with XL-VAE achieves significant improvements in translation accuracy over the direct translation and VQ-VAE baseline. 

In the future, we will enhance XL-VAE with domain adversarial training to better transfer the speech recognition ability from written languages to unwritten languages. We will test UWSpeech on more complicated sentences and language pairs. Furthermore, going beyond the proof-of-concept experiments in this work (we assumed English or Spanish is unwritten), we will apply UWSpeech on truly unwritten languages for speech to speech translation.

\section{Acknowledgments}
This work was supported by the Key Project of National Science Foundation of Zhejiang Province (No. LZ19F020002). This work was also partially funded by Microsoft Research Asia. Thanks are due to Shen Zhou for bringing strength during tough times.

\bibliography{st_unknown}
\end{document}


\linenumbers

\maketitle
\section{Datasets Details}
We choose Fisher Spanish-English dataset~\cite{post2013improved} for experiments, which contains telephone conversations of speech and text in Spanish and the corresponding text translations in English, with $130$K parallel training samples in total. Following~\cite{jia2019direct}, we synthesize the English speech according to the text using a commercialized text to speech system with a female speaker. We use the original training and development sets in the dataset. We only consider the most useful translation scenarios for unwritten languages (e.g. daily communication, travel translation, etc.) in which high-freq and simple words/sentences are usually used, so we obtain some common sentences from the full test set to form our test set by filtering the sentence with a threshold of word frequency. We set the threshold to $5$K for English and $10$K for Spanish and finally get about $1$K out of 3641 sentence pairs as the test set. We also conduct experiments on English to Spanish translation where the target Spanish speech is also synthesized using the commercialized text to speech system with another female speaker. The test set used in our paper is saved in CodeAndDataAppendix \textbf{\textit{UWSpeech\_testset}}.

We also provide the BLEU scores of our experiments in original Fisher test set in Table~\ref{tab_s2s_esen}.
\begin{table}[!h]
\centering
\small 
\begin{tabular}{c | c c c }
\toprule
Method & \textit{Direct Translation} & \textit{VQ-VAE} & \textit{UWSpeech} \\
\midrule
Speech to Speech & 0.80 & 3.42 & 9.35 \\
\bottomrule
\end{tabular}
\caption{The BLEU scores of Spanish to English speech to speech translation, where English is taken as the unwritten language.}
\label{tab_s2s_esen}
\vspace{-0.3cm}
\end{table}

For the written languages used in XL-VAE, we choose French, German and Chinese with speech and corresponding phoneme sequence. Both the German and French datasets are from Common Voice\footnote{https://voice.mozilla.org/}, where the German corpus contains about 280K training examples (325 hours) with 5007 different speakers and the French corpus contains 150K training examples (173 hours) with 3005 different speakers. For Chinese dataset, we use AIShell~\cite{aishell_2017} which contains about 140K training examples (178 hours) with 400 different speakers. We choose 320 hours, 160 hours and 160 hours from German, French and Chinese corpus respectively for training, and the others for development. We first convert the text in German, French and Chinese corpus into phoneme with our internal grapheme-to-phoneme conversion tool, and then map the phoneme to IPA~\cite{international1999handbook} according to our internal phoneme-to-IPA mapping table. We convert all the speech waveform in our experiments into mel-spectrogram
following~\cite{ren2019almost} with frame size of 50 ms and hop size of 12.5ms. 

\section{Model Configuration Details}
For the convolution and transposed convolution layers in the converter and inverter, the kernel size, stride size and filter size are set to $3$, $2$ and $256$ respectively. We stack $2$ convolution layers to set the down/up-sampling ratio $c$ to 4 according to the validation performance. We stack both $N=6$ layers of Transformer blocks in the converter and inverter, with hidden size of both the self-attention and feed-forward layers as $256$, and filter size of the feed-forward layer as $1024$. The size of the IPA dictionary $K$ is set as $177$ and the dimension of discrete token embeddings $D$ is set as $256$. We simply choose Griffin-Lim algorithm~\cite{griffin1984signal} as the vocoder to synthesize the final waveform of the speech. 

The translator performs speech to discrete tokens (IPA) translation, which follows the basic encoder-attention-decoder model structure in Transformer~\cite{vaswani2017attention}. The encoder has several additional convolution layers to transform the speech input, which follows the same configuration of the convolution layers in the converter (with a $1/4\times$ down-sampling ratio). The discrete token embedding size, hidden size, filter size, number of encoder and decoder layers of the translator are set to $256$, $256$, $1024$, $6$, $6$ respectively. 

\section{Pipeline Details and Codes}
\paragraph{Training Details}
We first train the converter, inverter and discrete token embeddings in XL-VAE. We up-sample the speech data of each written language (German, French, Chinese) to the same amount, and then up-sample the speech data of unwritten language (English or Spanish) to match to the total amount of written languages. We ensure there are an equal amount of data in written and unwritten languages in each mini-batch. We choose the $\lambda$ in Equation~\ref{eq_xlvae_loss} according to the validation performance and set $\lambda$ to $0.01$. The batch size is set to $25$K frames for each GPU and the XL-VAE training takes $200$K steps on $4$ Tesla V100 GPUs. 

After the training of XL-VAE, the phoneme error rates (PER) of three written languages (German, French and Chinese) on the development set are $16\%$, $21\%$ and $12\%$ respectively. We convert the target unwritten speech into discrete token sequence and keep the output discrete token sequence as it is, without removing any special or repeated tokens. We use the discrete token sequence generated by XL-VAE to train translator, with batch size of $16$K frames on each GPU and $100$K training steps on $4$ Tesla V100 GPUs. 

Our code is implemented based on tensor2tensor library~\cite{vaswani2018tensor2tensor}\footnote{https://github.com/tensorflow/tensor2tensor}. We attach the code in CodeAndDataAppendix \textbf{\textit{code}}, and the details of our code are explained in CodeAndDataAppendix \textbf{\textit{code/README.md}}

\paragraph{Inference and Evaluation}
During inference, we use the translator to generate discrete token sequence from source speech with beam search. We set beam size to $4$ and length penalty to $1.0$. We then directly use the inverter to transform the discrete token sequence back to target speech. 

To evaluate the accuracy of the speech translation, following the practice in~\cite{jia2019direct}, we pre-train an automatic speech recognition model (which can achieve 85.62 BLEU points on our test set and is comparable with~\cite{jia2019direct}) to generate the corresponding text of the translated speech, and then calculate the BLEU score~\cite{papineni2002bleu} between the generated text and the reference text. We report BLEU score using case insensitive BLEU with moses tokenizer\footnote{https://github.com/moses-smt/mosesdecoder/blob/master/ scripts/tokenizer/tokenizer.perl} and multi-bleu.perl\footnote{https://github.com/moses-smt/mosesdecoder/blob/master/ scripts/generic/multi-bleu.perl}. Due to the Fisher corpus has $4$ English references in the test set, we report 4-reference BLEU score for Spanish to English setting, and still report single-reference BLEU score for English to Spanish setting.

\section{Case Analyses and Demo Audios}
\begin{table}[!h]
\centering
\small 
\begin{tabular}{ l l }
\toprule
    Spanish (Source) & Yo no entendí lo que ella dijo. \\ 
    English (Target) & I didn't understand what she said. \\
    \midrule
    \textit{Direct Translation} & I don't know. \\
    \textit{VQ-VAE} & I didn't understand. \\
    \textit{UWSpeech} & I didn't understand what she say. \\
    \midrule
    \textit{IPA (Target)} & \textipa{ai ai n | d I g n n z E | 5 n Y s t t @ l 5 n n } \\
    &  \textipa{ t t | v O t | t i: s | E E: n } \\
    \textit{IPA (UWSpeech)} & \textipa{ai ai | d e n n n | a n n v y: s s t e n n n t } \\
    & \textipa{| v O 5 t | t d i: | E E l } \\
\bottomrule
\end{tabular}
\caption{Case 1}
\label{table_case1}
\end{table}

\begin{table}[!h]
\centering
\small 
\begin{tabular}{ l l }
\toprule
    Spanish (Source) & Qué tal, ¿de dónde eres? \\ 
    English (Target) & How's it going, where are you from? \\
    \midrule
    \textit{Direct Translation} & Had a price. \\
    \textit{VQ-VAE} & Like are you there are you from? \\
    \textit{UWSpeech} & How are you, where are you from? \\
    \midrule
    \textit{IPA (Target)} & \textipa{h h a: s | b I t t | Oy n n | O | K | j | v a: K m} \\
    \textit{IPA (UWSpeech)} & \textipa{h h au | 5 | j ø: | 5 | j e | v a: K m} \\
\bottomrule
\end{tabular}
\caption{Case 2}
\label{table_case2}
\end{table}

\begin{table}[!h]
\centering
\small 
\begin{tabular}{ l l }
\toprule
    Spanish (Source) & Yo soy puertoriqueña. \\ 
    English (Target) & I am Puerto Rican. \\
    \midrule
    \textit{Direct Translation} & Ah. \\
    \textit{VQ-VAE} & I'm from. \\
    \textit{UWSpeech} & I'm from Puerto Rico. \\
    \midrule
    \textit{IPA (Target)} & \textipa{ai n m | p o d d @ | v i: i: g e E n |} \\
    \textit{IPA (UWSpeech)} & \textipa{ai ai n n | f a ŋ | p o 5 d @ @ | v v e k k 5 | } \\
\bottomrule
\end{tabular}
\caption{Case 3}
\label{table_case3}
\end{table}

\begin{table}[!h]
\centering
\small 
\begin{tabular}{ l l }
\toprule
    Spanish (Source) & Halo, buenas noches. \\ 
    English (Target) & Hello good evening.  \\
    \midrule
    \textit{Direct Translation} & And a. \\
    \textit{VQ-VAE} & Hello video. \\
    \textit{UWSpeech} & Hello good evening. \\
    \midrule
    \textit{IPA (Target)} & \textipa{h a n l O | g l t t t | g I b n I I ŋ |} \\
    \textit{IPA (UWSpeech)} & \textipa{h a n l O | g l l d | i: i: v n ŋ ŋ} \\
\bottomrule
\end{tabular}
\caption{Case 4}
\label{table_case4}
\end{table}

All the corresponding audios are saved in MultimediaAppendix \textbf{\textit{cases}}.
For example, the UWSpeech audio of Case 1, shown in Table~\ref{table_case1} is named as $\left[case1\right]\left[UWSpeech\right]i\_didnt\_\\understand\_what\_she\_say.wav$.

\bibliography{st_unknown}
\bibliographystyle{aaai21}